\documentclass[pamm,a4paper,fleqn]{w-art}
\usepackage{times,cite,w-thm}
\usepackage[T1]{fontenc}
\usepackage[utf8]{inputenc}
\theoremstyle{plain}

\theoremstyle{definition}

\usepackage{graphicx}
\begin{document}

\TitleLanguage[EN]
\title[A spotlike edge state in plane Poiseuille flow]{A spotlike edge state in plane Poiseuille flow}


\author{\firstname{Stefan}  \lastname{Zammert}\inst{1,}%
  \footnote{Corresponding author: e-mail \ElectronicMail{Stefan.Zammert@gmail.com}}}

\address[\inst{1}]{\CountryCode[DE]Fachbereich Physik, Philipps-Universit\"at Marburg, Renthof 6, D-35032 Marburg}
\author{\firstname{Bruno} \lastname{Eckhardt}\inst{1,2,}}

\address[\inst{2}]{\CountryCode[NL] J.M. Burgerscentrum, Delft University of Technology, Mekelweg 2, 2628 CD Delft}
\AbstractLanguage[EN]
\begin{abstract}
We study the laminar turbulent boundary in plane Poiseuille flow at $Re=1400$ and $2180$ 
using the technique of edge tracking. For large computational domains the attracting state in the laminar-turbulent boundary  is
localized in spanwise and streamwise direction and chaotic.
\end{abstract}
\maketitle                   

\section{Introduction}
In many simple shear flows such as plane Couette flow, pipe flow, or plane Poiseuille flow (PPF)
a linearly stable laminar state and a turbulent state coexist. A convenient indicator that can be used
to explore which initial conditions are attracted to either state is the lifetime, i.e. the time it takes to return
to the laminar state. Close to the laminar state the lifetime is short and close to the turbulent state it is infinite if the state
is attracting and usually very long if the turbulent state is transient \cite{Schmiegel1997,Skufca2006}.
The distinct lifetimes in the two parts of the state space can be used to define an algorithm that tracks trajectories 
that are confined to the boundary between laminar and turbulent initial conditions.
This edge tracking algorithm \cite{Skufca2006,Schneider2007} uses a simple bisection in the amplitude of an 
initial condition to bracket a point on the laminar-turbulent boundary (LTB). The trajectories that evolve from the
two neighboring points will then approximate a trajectory in the LTB. With further bisections when the distance of 
the approximating trajectories becomes too large it is possible to follow trajectories on the LTB for very long times, until
they settle on an attracting set within the LTB, the so-called {edge states}. Edge states are useful from a 
fundamental point of view because they arise in subcritical bifurcations that are key to the turbulence transition, as
discussed in depth in \cite{Kreilos2012,Avila2013}.  They can also be used to find 
minimal disturbances that quickly become turbulent \cite{Duguet2013c,Cherubini14}. 

As an invariant state within the LTB, the edge state can come in many forms. In small computational domains in plane Couette it is a fixed point of the Navier-Stokes equation \cite{Schneider2008} 
and for narrow domains in PPF it is either a periodic orbit or a travelling wave, depending on the domain length   \cite{Zammert2014a,Zammert2014b}. In other cases, e.g in pipe flow, the edge states is chaotic \cite{Schneider2007}. We here 
show that the edge state becomes chaotic also for PPF in long and wide domains.

We define a Reynolds number $Re=U_{0}d/\nu$ for the system using the center-line velocity $U_{0}$ and half 
the distance between the plates. The laminar profile then is $U(y)=1-y^{2}$ and the plates are located at $y=\pm 1$. 
For our numerical simulations we use the \textit{Channelflow}-code \cite{Gibson2009b}.
The code solves the incompressible Navier-Stokes equations for a doubly-periodic domain with streamwise length 
$L_{x}$ and spanwise width $L_{z}$. For more details on the numerical methods and code verification we refer to the \textit{Channelflow}-manual (www.channelflow.org) and \cite{Zammert2014a,Zammert2014b}.

\section{The edge state in a large domain}
\begin{figure}
\centering
\includegraphics[]{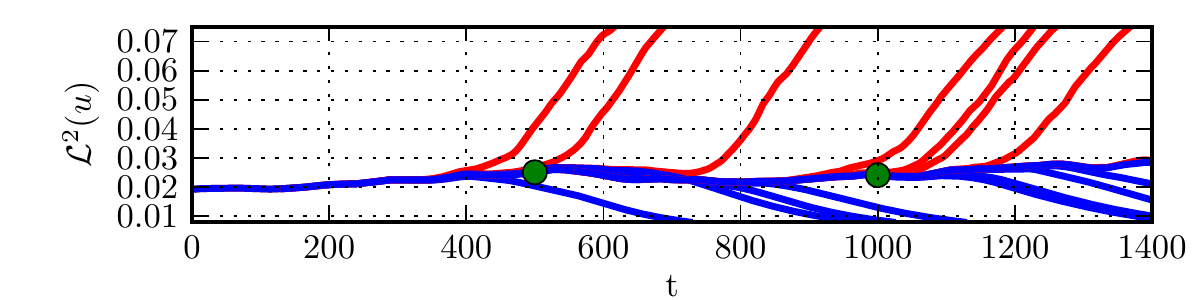}
\captionsetup{margin=4cc}
\caption{Edge tracking in a domain of size $L_{x} = 32\pi$ and $L_{z} = 8\pi$.
The trajectory on the laminar-turbulent boundary is approximated by trajectories on the turbulent (red) and laminar (blue) side.
Visualizations of the flowfields for the times marked by the green dots are shown in figure \ref{fig:2}. 
}
\label{fig:1}
\end{figure}

We performed edge tracking at $Re=1400$ in a computational domain with a width of $8\pi$ and a length of $32\pi$ 
with a numerical resolution of $N_{x}\times N_{y} \times N_{z}= 640 \times 65 \times 320$.
A picture of the edge tracking is shown in figure \ref{fig:1}.
The edge trajectory (approximated by the trajectories that become turbulent and laminar) 
does not converge to simple exact coherent state.

\begin{figure}
\centering
  \includegraphics[]{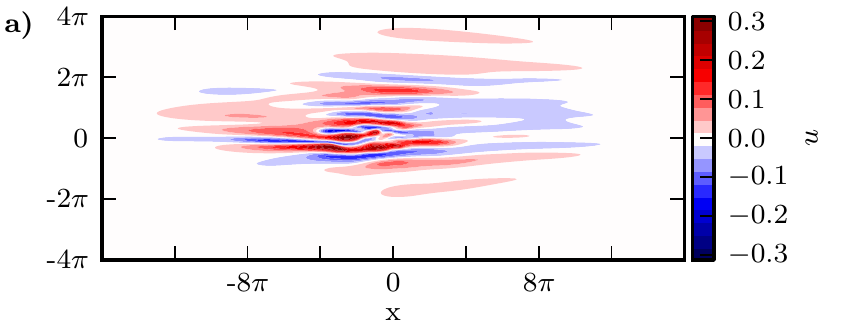}\includegraphics[]{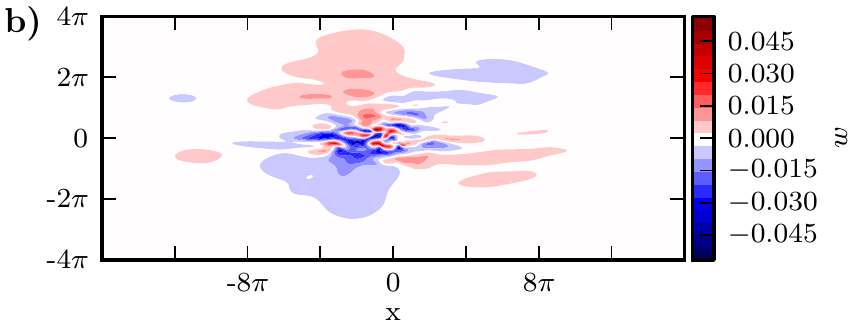}
  \includegraphics[]{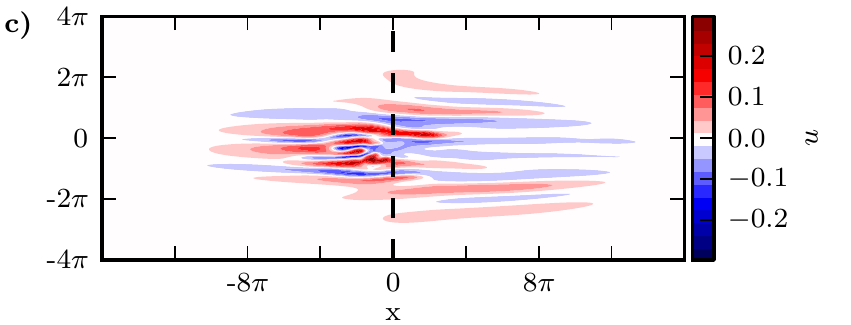}\includegraphics[]{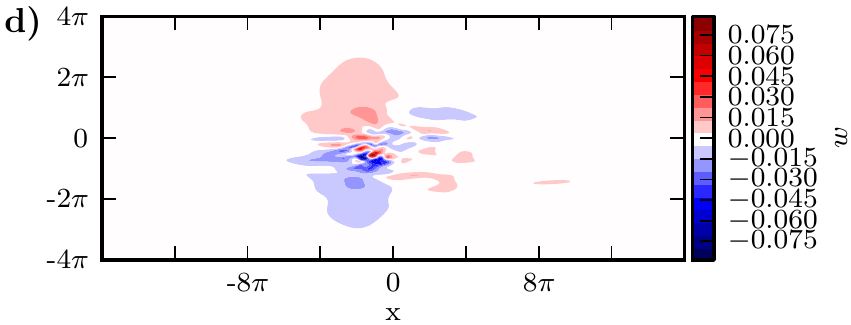}
  \includegraphics[]{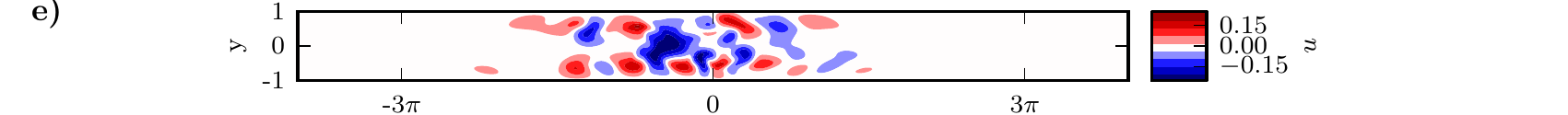}
\captionsetup{margin=4cc}
\caption{The panels a) and b) show the steamwise and spanwise velocity  in a plane parallel to the plates at $y=0.74$
for the flowfield on the edge trajectory at $t=500$. Panel c) and d) show the same visualizations for $t=1000$.
The direction of the flow is from left to right. For $t=1000$  the flow in a spanwise wall-normal cross section plane is shown in e). The streamwise position of this slice is marked in c) by a dashed line.}
\label{fig:2}
\end{figure}

For the times $t=500$ and $t=1000$ visualizations of the velocity in a plane parallel to the plates are shown in figure \ref{fig:2}.
In addition, for $t=1000$ at $x=0$ the streamwise velocity in the spanwise wall-normal plane is shown in figure \ref{fig:2}e).
It becomes evident that the flow structures are localized in streamwise and spanwise direction. 
For both times the flow structures are quite similar. The plots of the deviations of the streamwise velocity 
from the parabolic profile reveal streamwise streaks that are strongest in the center of the structure.
In the tail of the structure the streaks are aligned in streamwise direction while in the front the streaks are slightly oblique.
The cross-flow motion is localized to the center of the flow structure. 
The size of the flow structure varies only slightly over the observation time.
The overall flow structure resembles typical turbulent spots of PPF as described e.g. by Lemoutl et al. \cite{Lemoult2013}.
 
A further edge tracking at $Re=2180$ did also result in a spanwise and streamwise localized structure.
For this Reynolds number the length of the state is larger than for $Re=1400$ but the qualitatively features 
are unchanged.
\section{Conclusion}
While in computational domains that are either short and wide or long and narrow the edge states is an 
exact coherent state \cite{Zammert2014a,Zammert2014b} our simulations suggest that the edge state in large domains 
is a localized chaotic state that is quite similar to a turbulent spot.  The observation of a chaotic edge states 
in a spatial extended domains is quite similar to the case of plane Couette flow \cite{Schneider2008,Duguet2009} 
and pipe Poiseuille flow\cite{Mellibovsky2009}. 

\bibliographystyle{pamm}
\providecommand{\WileyBibTextsc}{}
\let\textsc\WileyBibTextsc
\providecommand{\othercit}{}
\providecommand{\jr}[1]{#1}
\providecommand{\etal}{~et~al.}

\end{document}